\documentclass[prd,twocolumn,groupedaddress,floatfix]{revtex4}    
\usepackage{graphics}                                    
\usepackage{color}     
\usepackage{amssymb}
\usepackage{mathrsfs}
\usepackage{float}
\usepackage{tabularx}
\usepackage{mathtools}
\usepackage{array}
\usepackage{multirow}
\def\nn{\nonumber}

\def\R2{{\mathcal R}^2}

\def\be{\begin{equation}}
\def\ee{\end{equation}}
\def\barr{\begin{array}{lr}}
\def\earr{\end{array}}
\def\bea{\begin{eqnarray}}
\def\eea{\end{eqnarray}}
\def\la{\langle}
\def\ra{\rangle}

\def\nn{\nonumber}

\def\d{\delta}
\def\e{\epsilon}

\def\o{\omega}

\def\s{\sigma}

\def\D{\Delta}

\def\G{\Gamma}

\begin{document}
\title{Minkowski Functionals for composite smooth random fields}
\author{Pravabati Chingangbam$^{1}$}
\email{prava@iiap.res.in}
\author{Fazlu Rahman$^{1,2}$}                               
\email{fazlu.rahman@iiap.res.in}
\affiliation{$^1$ Indian Institute of Astrophysics, Koramangala II Block, Bengaluru  560 034, India\\
$^2$ Department of Physics, Pondicherry University, R.V. Nagar, Kalapet, 605 014, Puducherry, India}


\begin{abstract}
 Minkowski functionals  quantify the morphology of smooth random fields.  They are widely used to probe statistical properties of cosmological fields. Analytic formulae for ensemble expectations of Minkowski functionals are well known for Gaussian and  mildly non-Gaussian fields.   In this paper we extend the formulae to composite fields which are sums of two fields and explicitly derive the expressions for the sum of uncorrelated mildly non-Gaussian and Gaussian fields. These formulae are applicable to   observed data which is usually a sum of the true signal and one or more secondary fields that can be either noise, or some residual contaminating signal. Our formulae provide explicit quantification of the effect of the secondary field on the morphology and statistical nature of the true signal. As examples, we apply the formulae to determine how the presence of Gaussian noise can bias the morphological properties and statistical nature of Gaussian and non-Gaussian CMB temperature maps.  
\end{abstract}

\maketitle

\section{Introduction}

 The morphology of smooth random fields~\cite{Adler:1981} that occur in nature contain a wealth of information about the physical processes that produce the fields. In the context of cosmology, various geometrical and topological statistical quantities have been proposed in the literature  to quantify the morphology of cosmological random fields and extract physical information. Of these, Minkowski functionals~\cite{Minkowski:1903,
 Gott:1990,Mecke:1994,Schmalzing:1997} are arguably the most widely used. 

There are $d+1$ Minkowski functionals on $d$ dimensional space. 
They completely characterize the translation and rotation invariant and additive  morphological properties  of excursion or level sets of smooth random field. 
Their functional shapes, as functions of the field levels, are determined by the nature of the field. Analytic expressions of the ensemble expectations of Minkowski
functionals for Gaussian random fields were derived by Tomita~\cite{Tomita:1986}.  Subsequently, analytic formulae for mildly non-Gaussian fields have been derived by Matsubara~\cite{Matsubara:2003,Matsubara:2011} as perturbative expansions in powers of the standard deviation of the field. In~\cite{Matsubara:2020} explicit formulae  for Minkowski functionals have been obtained upto second order of the standard deviation of the field in arbitrary $d$ dimensions.  An alternative approach to obtaining Minkowski functionals for non-Gaussian fields uses the Gram-Charlier expansion of the joint probability distribution of the field and its derivatives   \cite{Gay:2012PRD,Codis:2013MNRAS}.

Minkowski functionals belong to the wider class of morphological descriptors known as Minkowski tensors~\cite{McMullen:1997,Alesker:1999,Beisbart:2001gk,Hug:2008,Schroder2D:2009} which were introduced for random fields and cosmological applications in \cite{Chingangbam:2017uqv,Ganesan:2016,Appleby:2017uvb,Appleby:2018tzk}. Minkowski functionals can be obtained as the trace of the isometry preserving rank two Minkowski tensors. 

In practice, any observed data is a sum of the true signal and noise, and can also contain contamination by other signals. It is then of interest  to examine the impact of secondary fields (noise and/or other contaminating signals) on the morphological and statistical properties of the true signal in a quantifiable way. Towards this goal, in this paper we extend the formulae derived in  ~\cite{Matsubara:2020} to {\em composite fields}, by which we mean sums of two fields. We explicitly derive the expressions for the sum of uncorrelated mildly non-Gaussian and Gaussian fields. Though we focus here on fields on two dimensions the derivation can be generalized to higher dimensions.    
We then apply the formulae to two toy examples of composite fields. The first is the sum of a Gaussian temperature field of the cosmic microwave background (CMB) and Gaussian noise, while the second example is a sum of non-Gaussian
CMB and Gaussian noise fields. 

Minkowski functionals have been used to probe a plethora of physical properties of cosmological fields. 
These applications range from statistical properties of the CMB to probing the large scale structure of the universe and the epoch of reionization. Due to the large body of work on its applications we do not include all references here. A comprehensive list of applications and references can be found in~\cite{Matsubara:2020}. Our analysis here is motivated by our earlier work on understanding the statistical nature of Galactic foreground emissions~\cite{Rahman:2021azv,Rahman:2022}, impact of residual foreground contamination in the CMB~\cite{Chingangbam:2013}, and CMB polarization \cite{Ganesan:2014lqa,Chingangbam:2017PhLB}. Related investigations on Galactic foreground emissions have been carried out in~\cite{Rana:2018oft,Martire:2023,Duque:2024MNRAS}.

This paper is organized as follows. Section~\ref{sec:sec2} reviews the analytic formulae for ensemble expectations of Minkowski functionals for mildly non-Gaussian fields.  Section~\ref{sec:sec3} presents our extension of the formulae to composite fields. 
In section~\ref{sec:sec4}  as  application of the formulae we derive the bias introduced by noise on the morphology and non-Gaussianity of CMB temperature maps. We end with a summary and discussion of our results in section~\ref{sec:sec5}.  

\section{Review of Minkowski functionals for a mildly non-Gaussian field}
\label{sec:sec2}

Let $f$ be an isotropic smooth random field on a general $d$ dimensional space   $\mathcal{M}$. 
The set 
$Q_{\nu}=\big\{x\in\mathcal{M}|f(x)\ge\nu\s_0\big\}$ 
is called the excursion or level set indexed by $\nu$. Let the iso-field boundaries of $Q_{\nu}$ be denoted by $\partial Q_{\nu}$. Then the Minkowski functionals (MFs) are functionals of $f$ that quantify the morphology of each $Q_{\nu}$. 

In $d$ dimensional space there are $d+1$ MFs. 
In the following we review the analytic expressions for ensemble expectations of the MFs  for mildly non-Gaussian fields in terms of the spectral properties of the fields. All the expressions in this section follow~\cite{Matsubara:2020}. We assume the mean of the field to be zero so as to simplify the discussion, and it is straightforward to generalize to fields with non-zero means. 

Let us define the following spectral parameters for a given field $f$,
\be
\s_0 ^2 \equiv \la f^2 \ra, \quad \s_1^2 \equiv \la |\nabla f|^2\ra, \quad \s_2^2 \equiv \la (\nabla^2 f)^2\ra,
\ee
and the quantity $r_c$,
\be
\quad r_c \equiv \s_0 /\s_1,
\ee
which physically represents a typical length scale that quantifies the {\em typical spatial size} of structures for an isotropic field. 
For non-Gaussian $f$, the higher order connected cumulants are non-zero. 
The generalized skewness cumulants are defined as 
\bea
S^{(0)} & =& \frac{\la f^{3}\ra_{c}}{\sigma_{0}^{4}}, \nn \quad \\
S^{(1)} &=&\frac{3}{2}\frac{\la f|\nabla f|^{2}\ra_{c}}{\sigma_{0}^{2}\sigma_{1}^{2}},\nn\\
S^{(2)} &= & \frac{-3d}{2(d-1)}\frac{\la |\nabla f|^{2}\nabla^{2} f\ra_{c}}{\sigma_{1}^{4}}. \label{eq:Skew}
\eea

\pagebreak
The generalized kurtosis cumulants are
\bea
K^{(0)} \hspace{-.2cm}&=& \frac{\la f^{4}\ra_{c}}{\sigma_{0}^{6}}, \quad \nn\\ 
K^{(1)} \hspace{-.2cm}&=& 2\frac{\la 2f^{2}|\nabla f|^{2}\ra_{c}}{\sigma_{0}^{4}\sigma_{1}^{2}}, \nn\\
K^{(2)}_{1}\hspace{-.2cm}&=&\frac{-2d}{(d+2)(d-1)}\frac{(d+2)\la f|\nabla f|^{2}\nabla^{2} f\ra_{c}+\la|\nabla f|^{4}\ra_{c}}{\sigma_{0}^{2}\sigma_{1}^{4}}, \nn\\  
K^{(2)}_{2}\hspace{-.2cm}&=&\frac{-2d}{(d+2)(d-1)}\frac{(d+2)\la f|\nabla f|^{2}\nabla^{2} f\ra_{c}+d\la|\nabla f|^{4}\ra_{c}}{\sigma_{0}^{2}\sigma_{1}^{4}}, \nn \\ 
K^{(3)}\hspace{-.2cm}&=& \frac{2d^{2}}{(d-1)(d-2)}\frac{\la |\nabla f|^{2}(\nabla^{2} f)^{2}\ra_{c}-\la |\nabla f|^{2}f_{ij}f_{ij}\ra_{c}}{\sigma_{1}^{6}}.\nn\\ \label{eq:Kurt}
&& 
\eea
The subscript $c$ on the the angle brackets indicate that these quantities are  connected cumulants. As the field is mean-free, the third-order cumulants are equal to the third-order moments. Note that   
$K_1^{(1)}$ and $K_2^{(2)}$ diverge for $d=1$. Their difference, however, is finite and that is what enters in the formulae for MFs. $K^{(3)}$ is divergent for $d=1,2$ but that is not a problem since it enters the formulae for MFs only for $d\ge 3$. 

The fourth-order cumulants are given in terms of the moments 
as,
{\small{
\begin{eqnarray}
\langle f^{4} \rangle_{c}&=& \langle f^{4} \rangle -3\sigma_{0}^{4},\nn \\
\langle f^{2}|\nabla f|^{2}\rangle_{c}&=& \langle f^{2}|\nabla f|^{2}\rangle -\sigma_{0}^{2}\sigma_{1}^{2},\nn\\
\langle f|\nabla f|^{2} \nabla^{2} f\rangle_{c} &=& \langle f|\nabla f|^{2} \nabla^{2} f\rangle +\sigma_{1}^{4}, \nn \\
\langle |\nabla f|^{4} \rangle_{c} &=& \langle |\nabla f|^{4} \rangle -\frac{d+2}{2}\sigma_{1}^{4},\nn \\
\langle |\nabla f|^{2}(\nabla^{2}f)^{2},\rangle_{c}&=&\langle |\nabla f|^{2}(\nabla^{2}f)^{2}\rangle-\sigma_{1}^{2}\sigma_{2}^{2}, \nn\\
\langle |\nabla f|^{2}f_{ij}f_{ij}\rangle_{c}&=&\langle |\nabla f|^{2}f_{ij}f_{ij}\rangle-\sigma_{1}^{2}\sigma_{2}^{2}. 
\end{eqnarray}
}}
Here $f_{ij}$ are derivatives with respect to $i^{\rm th}$ and $j^{\rm th}$ coordinates of ${\cal M}$, with $i,j=1,2,...,d$.  

Keeping upto $\s_0^2$ order, the expressions for the ensemble expectation of MFs per unit volume for mildly non-Gaussian fields are given by 
\begin{widetext}
\bea
\bar{V}_{k}^{(d)}(\nu)&\simeq& 
A_k e^{-\nu^{2}/2} \Bigg[ H_{k-1}(\nu)+\left\{\frac{1}{6}S^{(0)}H_{k+2}(\nu)+\frac{k}{3}S^{(1)}H_{k}(\nu)+\frac{k(k-1)}{6}S^{(2)}H_{k-2}(\nu)\right\} \sigma_{0} \nn \\
&& +\Bigg\{\frac{1}{72}(S^{(0)})^{2}H_{k+5}(\nu)+\Bigg(\frac{1}{24}K^{(0)}  +\frac{k}{18}S^{(0)}S^{(1)}\Bigg)H_{k+3}(\nu) \nn \\
&& +k\Bigg( \frac{1}{8}K^{(1)} +\frac{k-1}{36}S^{(0)}S^{(2)}+\frac{k-2}{18}(S^{(1)})^{2}\Bigg)H_{k+1}(\nu)\nn \\
&& +k\Bigg(\frac{k-2}{16}K_{1}^{(2)}+\frac{k}{16}K_{2}^{(2)} +\frac{(k-1)(k-4)}{18}S^{(1)}S^{(2)}\Bigg) H_{k-1}(\nu) \nn \\
&& + k(k-1)(k-2)\Bigg(\frac{1}{24}K^{(3)}+ \frac{k-7}{72}(S^{(2)})^{2}\Bigg)
H_{k-3}(\nu)\Bigg\}\sigma_{0}^2+\mathcal{O}(\sigma_{0}^{3})\Bigg],
\label{eqn:fullMFs}
\eea
\end{widetext}
where $k=0,1,\ldots,d$. $H_k(\nu)$ are the (probabilist) Hermite polynomials, and  $H_{-1}(\nu)= \sqrt{\frac{\pi}{2}}e^{\nu^2/2}$. 
The amplitude $A_k$ is given by
\be
A_{k}=\frac{1}{(2\pi)^{(k+1)/2}} \frac{\o_d}{\o_{d-k}\o_k} \left(\frac{\s_1}{\sqrt{d}\s_0} \right)^k. 
\ee
The factors $\o_n$ for integer $n\ge 0$ are given by $\o_n=\pi^{n/2}/\G(n/2+1)$. 
So, we have  $\omega_0=1,\ \omega_1=2,\ \omega_2= \pi$, $\o_3=4\pi/3$, and so on. 

\section{Minkowski functionals for composite  fields in two dimensions}
\label{sec:sec3}

In two dimensions (2D) the excursion set boundaries $\partial Q_{\nu}$ form closed contours. There are three MFs. The first one,  $V_0$, gives the area fraction of $Q_{\nu}$, while the second one, $V_1$, gives the contour length per unit volume.  The third MF, $V_2$, is the integral of the geodesic curvature over  $\partial Q_{\nu}$. In flat 2D space $V_2$ is equal to the difference between the numbers of connected regions and holes, usually referred to as the Euler characteristic, per unit volume.  
On curved 2D space, however, $V_2$ is no longer equal to the Euler characteristic, and is not a topological quantity.  As a curvature integral it  still captures the morphological and statistical properties of the field. 

For a Gaussian field,   
the ensemble expectation of the MFs given by  Eq.~\ref{eqn:fullMFs} reduce to the simple form,
\begin{equation}
V_{k}^G(\nu)=A_k\,e^{-\nu^{2}/2}v_k^G(\nu),
\label{eqn:gmf}
\end{equation}
where $k=0,1,2$. The coefficients $A_k$ are,
\be
A_0=\frac{1}{\sqrt{2\pi}},\  A_1=\frac{1}{8\sqrt{2}}\frac{\s_1}{\s_0},\ A_2=\frac{1}{4\sqrt{2}\,\pi^{3/2}}\left(\frac{\s_1}{\s_0}\right)^2,
\label{eqn:Ak2}
\ee
and the functions $v_0^{\rm (G)}$ are
\begin{eqnarray}
 v_0^{\rm (G)} = \sqrt{\frac{\pi}{2}}\,e^{\nu^{2}/2}\,\text{erfc}\left(\frac{\nu}{\sqrt{2}}\right),\  
v^{(\rm G)}_1 = 1,\ 
v^{(\rm G)}_2(\nu) = \nu. \nn \\
\label{eqn:gvk}
\end{eqnarray}

For a mildly non-Gaussian field 
the ensemble expectation of the MFs are given by,  
\begin{equation}
V_{k}(\nu)=A_k\,e^{-\nu^{2}/2}v_k(\nu),
\label{eqn:mf}
\end{equation}
where the functions $v_k$ are given by  
\begin{equation}
v_{k}=v_{k}^{(\rm G)}+v_{k}^{(1)}\sigma_0+v_{k}^{(2)}\sigma_0^{{2}}+\mathcal{O}(\sigma_0^{3}). 
\end{equation} 
The first-order non-Gaussian terms are given in terms of the skewness cumulants, as,
\begin{eqnarray}
v_{0}^{(1)}&=&\frac{S^{(0)}}{6}H_{2}(\nu),\\
v_{1}^{(1)}&=&\frac{S^{(0)}}{6}H_{3}(\nu)+\frac{S^{(1)}}{3}H_{1}(\nu),\\
v_{2}^{(1)}&=&\frac{S^{(0)}}{6}H_{4}(\nu)+\frac{2S^{(1)}}{3}H_{2}(\nu)+\frac{S^{(2)}}{3}H_{0}(\nu).
\end{eqnarray}
The second-order non-Gaussian terms are given in terms of kurtosis cumulants,   as,
\begin{eqnarray}
  v_{0}^{(2)}(\nu)&=&\frac{\left(S^{(0)}\right)^2}{72}H_{5}(\nu)+\frac{K^{(0)}}{24}H_{3}(\nu), \\
v_{1}^{(2)}(\nu)&=&\frac{\left(S^{(0)}\right)^2}{72}H_{6}(\nu) +\frac{1}{24}\left(K^{(0)}+\frac{4}{3}S^{(0)}S^{(1)}\right)H_4(\nu) \nn \\
&&   +\frac{1}{8}  \left(K^{(1)}+\frac{4}{9}\left(S^{(1)}\right)^2\right)  H_2(\nu)\nn \\
&& -\frac{1}{16}\left(K^{(2)}_{1}-K^{(2)}_{2}\right)H_0(\nu), \\
   v_{2}^{(2)}(\nu)&=&\frac{\left(S^{(0)}\right)^2  }{72}H_{7}(\nu)+\frac{1}{24}\left(K^{(0)}+\frac{8}{3}S^{(0)}S^{(1)}\right)H_{5}(\nu) \nn \\
   && +\frac{1}{4}\left(K^{(1)}+\frac{2}{9}S^{(0)}S^{(2)}\right)H_{3}(\nu) \nn \\
   && +\frac{1}{4}\left(K^{(2)}-\frac{8}{9}  S^{(1)}S^{(2)}\right)H_{1}(\nu). 
\end{eqnarray}
We will find it useful to introduce the analytic forms of the non-Gaussian deviations of the MFs:
\bea
\Delta V_{k}^{\rm ana} &=& V_{k} -V_{k}^{G}, \nn\\
&\simeq&     A_k\,e^{-\nu^{2}/2}\left( v_k^{(1)}\s_0 + v_k^{(2)}\s_0 \right).
   \label{eqn:DVk_ana}
\eea

Let us now consider the field to be given by $f=u+v$, where $u$ and $v$ are either Gaussian or mildly non-Gaussian smooth random fields.   The MFs of $f$ will have contributions from the individual MFs of $u$ and $v$. In general these contributions will not simply add up or get averaged because MFs are integral geometric  quantities. 
Our aim in this section is to express the formulae for the MFs of $f$ in terms of MFs of the field $u$, and determine the additional terms and factors introduced by the presence of $v$. 
For physical applications we consider $u$ to be the {\em signal} of interest, while $v$ is  either noise or a contaminating field. 

The expressions given till Eq.~\ref{eqn:DVk_ana} hold for any given single field.  In what follows, for clarity of notation we will use  superscripts  `$f$' ,`$u$' or `$v$' appropriately for the quantities $\s_0,\s_1,r_c,S^{(i)},K^{(i)}$, and other quantities constructed from these, so as to specify the field under consideration.  

To keep the discussion general, in addition to the spectral parameters and the typical size of structures of $u,v,f$, let us also introduce the cross correlations of $u$ and $v$, and of their first derivatives, which we denote by,
\be
c^{uv} = \frac{\la uv\ra}{\s_0^u \s_0^v}, \quad 
c_1^{uv} = \frac{\la\nabla   u \cdot \nabla v\ra}{\s_1^u\s_1^v}. 
\ee
In order to compare the spectral parameters of the two fields, let us introduce the following two parameters,
\bea
\epsilon &\equiv& \frac{\s_0^v}{\s_0^u}, \quad p \equiv \frac{r_c^u}{r_c^v} = \e^{-1}\frac{\s_1^v}{\s_1^u}. 
\eea
 $\e$ compares the size of fluctuations of the field values of $u$ and $v$. If  $u$ is the physical signal of interest and $v$ is a noise field, then $\e$ is the inverse of the signal-to-noise ratio (SNR) of the two fields.  $p$ compares the size of {\em spatial fluctuations} of $u$ and $v$.  By definition we have the following ranges of the four parameters $\e, p, c^{uv}$ and $c_1^{u,v}$ to be
\be
0 < \e< \infty,\quad 0<p<\infty, \quad |c^{uv}| \le 1, \quad |c_1^{uv}| \le 1.
\ee 
For fields in 2D the four parameters $\e, p, c^{uv}, c_1^{uv}$ determine the relative importance of the fields $u$ and $v$ in the MFs of their composite field. Note that  
for fields in dimensions higher than two, we need additional parameters that compare the  spectral properties of the second derivatives of the two fields.

To express $A_k$, and the skewness and kurtosis cumulants in terms of $\e,\,p,\,c^{uv}$ and $c_1^{uv}$  we need the following expressions, 
\bea
{\left(\s_0^f\right)^2} & = &  {\left(\s_0^u\right)^2  \left(1+\e^2+2\e c^{uv}\right)}, \label{eqn:s0}\\ 
{\left(\s_1^f\right)^2} & = &  {\left(\s_1^u\right)^2} {\left(1+\e^2p^2+2\e p c_1^{uv}\right)}.\label{eqn:s1}  
\eea
Below we derive one by one the expressions for amplitude and skewness and kurtosis terms of the MFs of $f$ in terms of contributions from $u$ and $v$.


\subsection{Amplitude of MFs}
\label{sec:sec3.1}

Since the amplitude $A_k^f$ of the MFs, for $k=1,2$, are  proportional to $(r_c^f)^{-k}$, we  need to express $r_c^f$  in terms of $r_c^u$ and the parameters $\e,p,c^{uv}$ and $c_1^{uv}$. We get,
\bea
\left(r_c^f\right)^{-k} &=& 
=  \left(r_c^u\right)^{-k}\left[ \frac{1+\e^2 p^2 +2pc_1^{uv}}{1+\e^2+2\e c^{uv}} 
\right]^{k/2}. \label{eqn:rcf}
\eea
For $k=1$ we can take the positive square root of the right hand side. 
By inserting the above in Eq.~\ref{eqn:mf} we obtain the amplitude of the MFs of the composite field, with contributions from the two fields explicitly factored out. 

Note that the sum of two Gaussian uncorrelated fields is also a Gaussian field. Hence, if $u$ and $v$ are uncorrelated, this amplitude change will be the only effect of the secondary field and the shape of the MFs will be as given by the Gaussian expectations.

\begin{figure}[t]
\begin{center}
\includegraphics[scale=0.6]{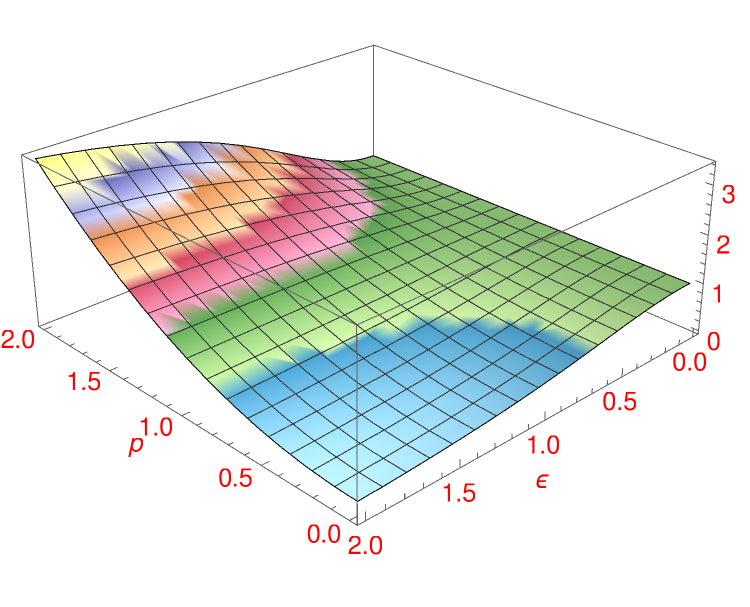}
\end{center}
\vspace{-0.6cm}
\caption{Plot of the factor $B(\e,p)=(1+\e^2p^2)/(1+\e^2)$ that relates $\left(r_c^f\right)^{-2}$ and $\left(r_c^u\right)^{-2}$ in Eq.~\ref{eqn:rcf}. The green regions are where $B(\e,p)$ has values $\sim 1$. Blue regions are where it has values less than one, while the other colour bands towards the top left indicate regions where it has values larger than one. } 
\label{fig:Afact}
\end{figure}

Let us now examine some special cases. Let us consider $u$ and $v$ to be uncorrelated, and so  we have $c^{uv} =0$ and $ c_1^{uv}=0$. Then Eq.~\ref{eqn:rcf} simplifies to
\bea
\left(r_c^f\right)^{-k} &=&  \left(r_c^u\right)^{-k}\left[ \frac{1+\e^2 p^2}{1+\e^2} \right]^{k/2}. \label{eqn:Ak_uncorr}
\eea
In the limit $\e\ll 1$ and $p\ll 1$, we get $\left(r_c^f\right)^{-2} \simeq  \left(r_c^u\right)^{-2}$, which is as expected. This is the limiting case when the standard deviation of $v$ is much lower than that of $u$, and the typical size of its structures is much larger than that of $u$.   

For practical applications it is of interest to examine values of $\e$ and $p$ in the vicinity of one. 
For this it is instructive to visualize the factor  $B(\e,p)=(1+\e^2 p^2)/(1+\e^2)$ that relates the spatial sizes of two fields in Eq.~\ref{eqn:rcf}. A plot of $B(\e,p)$ is shown in Fig.~\ref{fig:Afact}.  It is interesting to note the  following different cases based on the values of $\e$ and $p$.
\begin{enumerate}
\item If $\e=1$, with $p$ unconstrained, then we have 
\bea
\left(r_c^f\right)^{-2} &=&  
\frac12\left(\frac{1}{\left(r_c^u\right)^2 }  + \frac{1}{\left(r_c^v\right)^2}  \right)  . 
\eea
This implies that $A_2^f$ is the average of the corresponding amplitudes of the two fields. $A_1^f$ is, however, not the corresponding average.  

\item If $p=1$, with $\e$ unconstrained, then trivially $r_c^f = r_c^u = r_c^v$.  
This tells us that as long as the two fields $u$ and $v$,  normalized by their respective standard deviations, have the same spatial size of structures, then $f$ will also have the same size of structures, regardless of the value of $\e$.  So, the amplitudes of  all the MFs will be the same.
\item If $p<1$, with $\e$ unconstrained, then we get $\left(r_c^f\right)^{-2} < \left(r_c^u\right)^{-2}$. In this case the amplitudes of the MFs of $f$ will be decreased relative to that of $u$ field. 

For $\e<1$, expanding the denominator to $\e^2$ order, we get
\bea
\left(r_c^f\right)^{-2} 
&\simeq&   \left(r_c^u\right)^{-2}\left[ 1-\e^2(1-p^2)\right],
\eea
while for $\e>1$ we have 
 \bea
 \left(r_c^f\right)^{-2} &\simeq&  \left(r_c^u\right)^{-2} \left( 1/\e^2+p^2\right). 
 \eea

\item If $p>1$ with $\e$ unconstrained, then we get $\left(r_c^f\right)^{-2} > \left(r_c^u\right)^{-2}$. In this case the amplitudes of the MFs of $f$ will be increased relative to that of $u$ field. 

For $\e<1$ we get
 \bea
 \left(r_c^f\right)^{-2} &\simeq&  \left(r_c^u\right)^{-2} \left[1+\e^2p^2\right],
 \eea
while for $\e>1$ we get
\bea
 \left(r_c^f\right)^{-2} &\simeq&  \left(r_c^u\right)^{-2} p^2 =\left(r_c^v\right)^{-2}.
 \eea
\end{enumerate}

These cases inform us that the contribution of $v$  to the amplitudes of the MFs of the composite field will be small only if  $p\sim 1$, or $\e\to 0$ and $\e p\to 0$. 

 For the general situation where $u$ and $v$ are correlated, the amplitudes of the MFs of $f$ relative to $u$, will be determined by  whether the factor $(1+\e^2 p^2 +2pc_1^{uv})/(1+e^2+2\e c^{uv})$ is equal to, greater, or less than one. Positive correlation with $c^{uv} > 0$ will tend to decrease the amplitude, and vice versa. On the other hand, positive correlation of the first derivatives, $c_1^{uv}>0$, will tend to increase the amplitude, and vice versa.

\subsection{Generalized skewness and kurtosis cumulants}
\label{sec:sec3.2}

Next, we examine the expressions for the generalized skewness and kurtosis cumulants of the composite field in terms of the  cumulants of $u$ and parameters $\e$ and $p$. To simplify  the discussion we again consider $u,v$ to be uncorrelated. 

Let $u$ be mildly non-Gaussian and $v$ be  Gaussian.  The non-Gaussian deviations of the MFs of $f$ will be inherited from $u$. The generalized skewness cumulants of $f$ expressed in terms of the corresponding cumulants of $u$ and $\e,p$ are 
\bea
S^{(0)\,f}\s_0^f &=&  S^{(0)\,u}\s_0^u  \ \frac{1}{\left(1+\e^2\right)^{3/2}} , \label{eqn:S0f}\\ 
S^{(1)\,f}\s_0^f &=&  S^{(1)\,u}\s_0^u  \ \frac{1}{\left(1+\e^2\right)^{1/2}  \left(1+\e^2p^2\right)} , \label{eqn:S1f} \\ 
S^{(2)\,f}\s_0^f &=& S^{(2)\,u}\s_0^u  \ \frac{\left(1+\e^2\right)^{1/2} }{ \left(1+\e^2 p^2\right)^2 }. \label{eqn:S2f}
\eea
The factors containing $\e$ and $p$ on the right hand sides of the above equations are different for different skewness cumulants. This difference  implies that the presence of $v$ will change the relative strengths of the these cumulants, which is tantamount to changing the nature of non-Gaussianity of $f$ compared to $u$. Eqs.~\ref{eqn:S0f} and \ref{eqn:S1f} tell us that  $|S^{(0)f}|\s_0^f < |S^{(0)u}|\s_0^u $ and $|S^{(1)f}|\s_0^f< |S^{(0)u}|\s_0^u $ for all values of $\e,p$. On the other hand, $|S^{(2)f}|\s_0^f$ can be smaller or larger than $|S^{(2)u}|\s_0^u$, depending on the values of $\e,p$.  

The generalized kurtosis cumulants are,
\bea
K^{(0)\,f}(\s_0^f)^2 
&=& K^{(0)\,u}\left(\s_0^u\right)^2 \frac{1}{\left(1+\e^2\right)^{2} } - \frac{6\e^2}{(1+\e^2)^3},\label{eqn:K0f}\\ 
K^{(1)\,f}(\s_0^f)^2 
&=& K^{(1)\,u}\left(\s_0^u\right)^2 \frac{1}{\left(1+\e^2\right)  \left(1+\epsilon^{2}p^{2}\right)}  \nn\\
&& -\frac{3\e^2+3\e^2p^2+8\e^4p^2}{(1+\e^2)(1+\e^2p^2)},\label{eqn:K1f}\\  
K_2^{(1)\,f} (\s_0^f)^2 
&=& K_2^{(1)\,u}\left(\s_0^u\right)^2  \frac{1}{\left(1+\epsilon^{2}p^{2}\right)^2 } 
 + \frac{\e^2p^2}{(1+\e^2p^2)^2}, \nn \\ \label{eqn:K21f}\\
K_2^{(2)\,f} (\s_0^f)^2 
&=& K_2^{(2)\,u}\left(\s_0^u\right)^2  \frac{1}{\left(1+\epsilon^{2}p^{2}\right)^2 }
 +2 \frac{\e^2p^2}{(1+\e^2p^2)^2}. \nn\\ \label{eqn:K22f}
\eea
These equations again indicate that the presence of field $v$ will modify the nature of non-Gaussianity of $f$ compared to $u$ in a highly non-trivial manner. The kurtosis cumulant $K^{(3)}$ is zero in this case. It will be non-zero on spaces of dimension higher than two.

By inserting the right hand side of Eq.~\ref{eqn:rcf} for the amplitude, and Eqs.~\ref{eqn:S0f}  to \ref{eqn:K22f} for the cumulants in Eq.~\ref{eqn:mf}, we obtain the MFs of the composite field with contributions from the two constituent fields explicitly factored out. 
Corresponding expressions for the generalized skewness and kurtosis cumulants for Gaussian $u$ and non-Gaussian $v$, and when both fields are mildly non-Gaussian can be worked out in a similar manner. For the most general case when the two component fields are correlated, there will also be additional terms proportional to the cross-correlations between the two fields and their first and second derivatives. 
We do not include these cases here.

\section{Applications to examples of composite fields}  
\label{sec:sec4}

We now consider toy examples of composite fields. As mentioned in section~\ref{sec:sec3} we consider $f$ to be a composite of a signal field $u$ and an secondary field $v$ which may be noise or some other contaminating field. The purpose of this section is to quantitatively show the bias introduced by the secondary field on the MFs of the signal field using the analytic formulae obtained in section~\ref{sec:sec3}, and to establish their agreement with numerical calculations of the MFs.   We use the method introduced in~\cite{Schmalzing:1998,Schmalzing:1997} for numerical calculations of MFs. An alternative method of calculation can be found in \cite{Ducout:2013}.

\subsection{Sum of two uncorrelated Gaussian fields: CMB temperature and noise}
\label{sec:4.1}

\begin{figure}[t]
\begin{center}
 \includegraphics[scale=0.29]{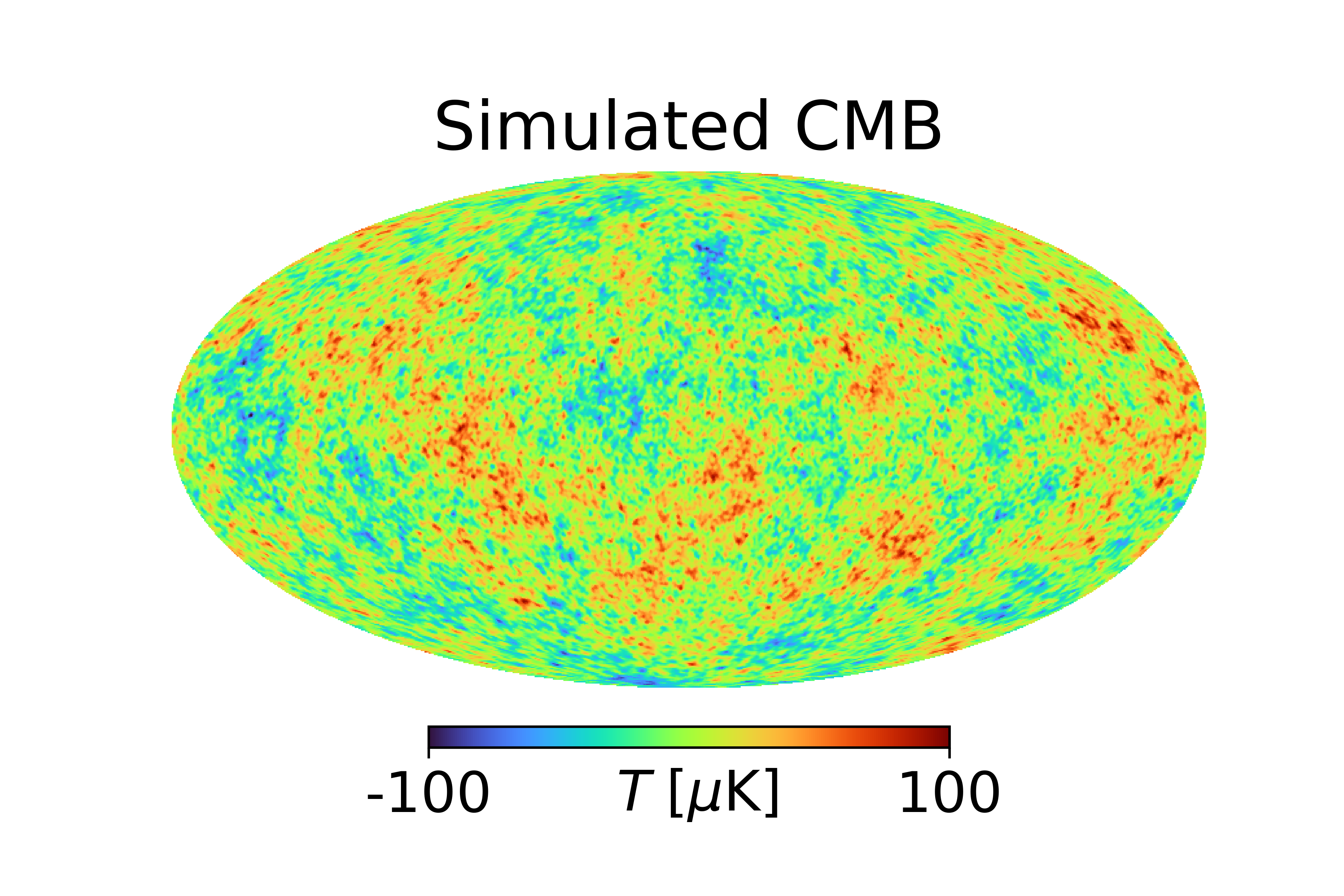}
 \hspace{-.56cm}\includegraphics[scale=0.29]{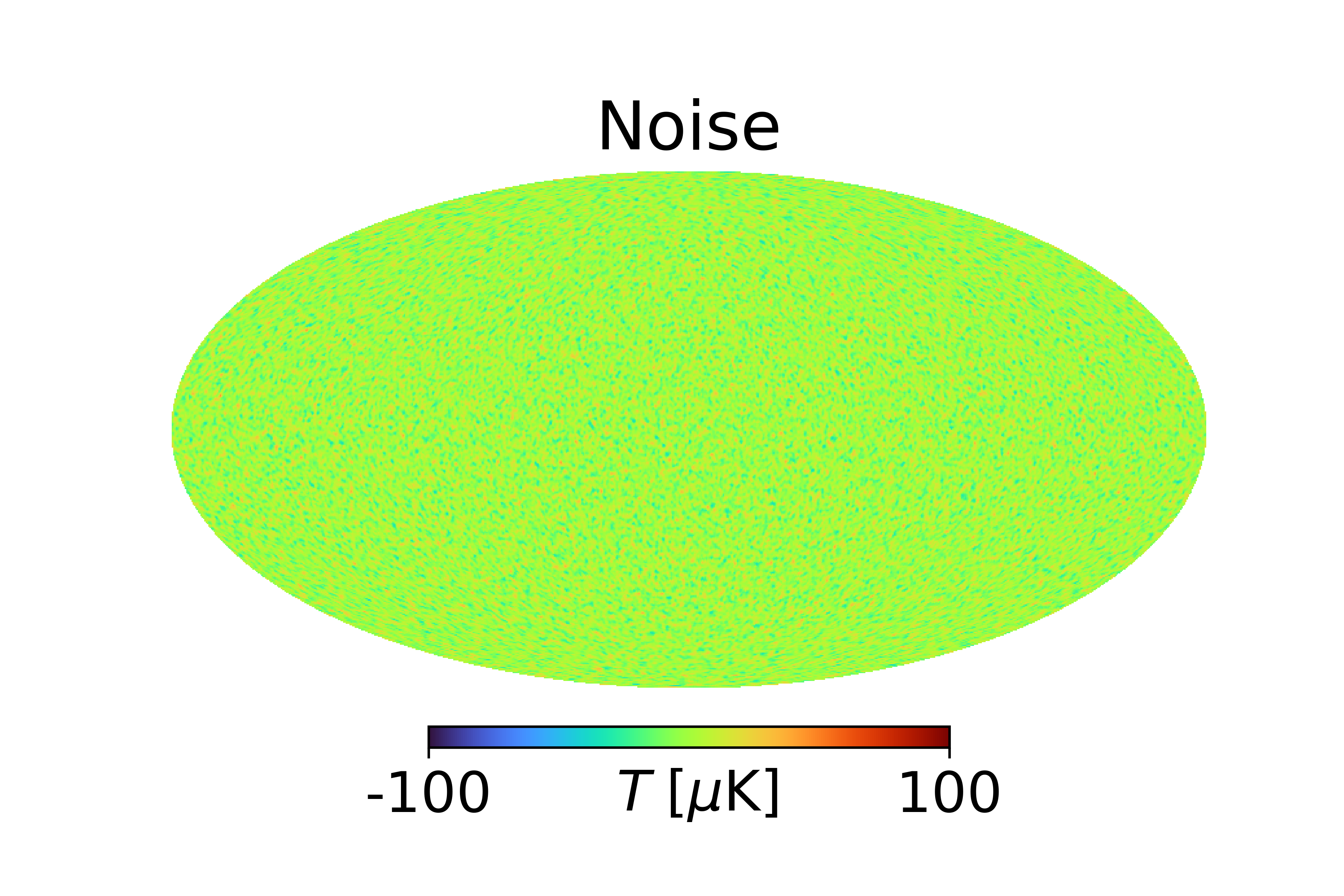}\\
  \vspace{-.3cm}
 \includegraphics[scale=0.29]{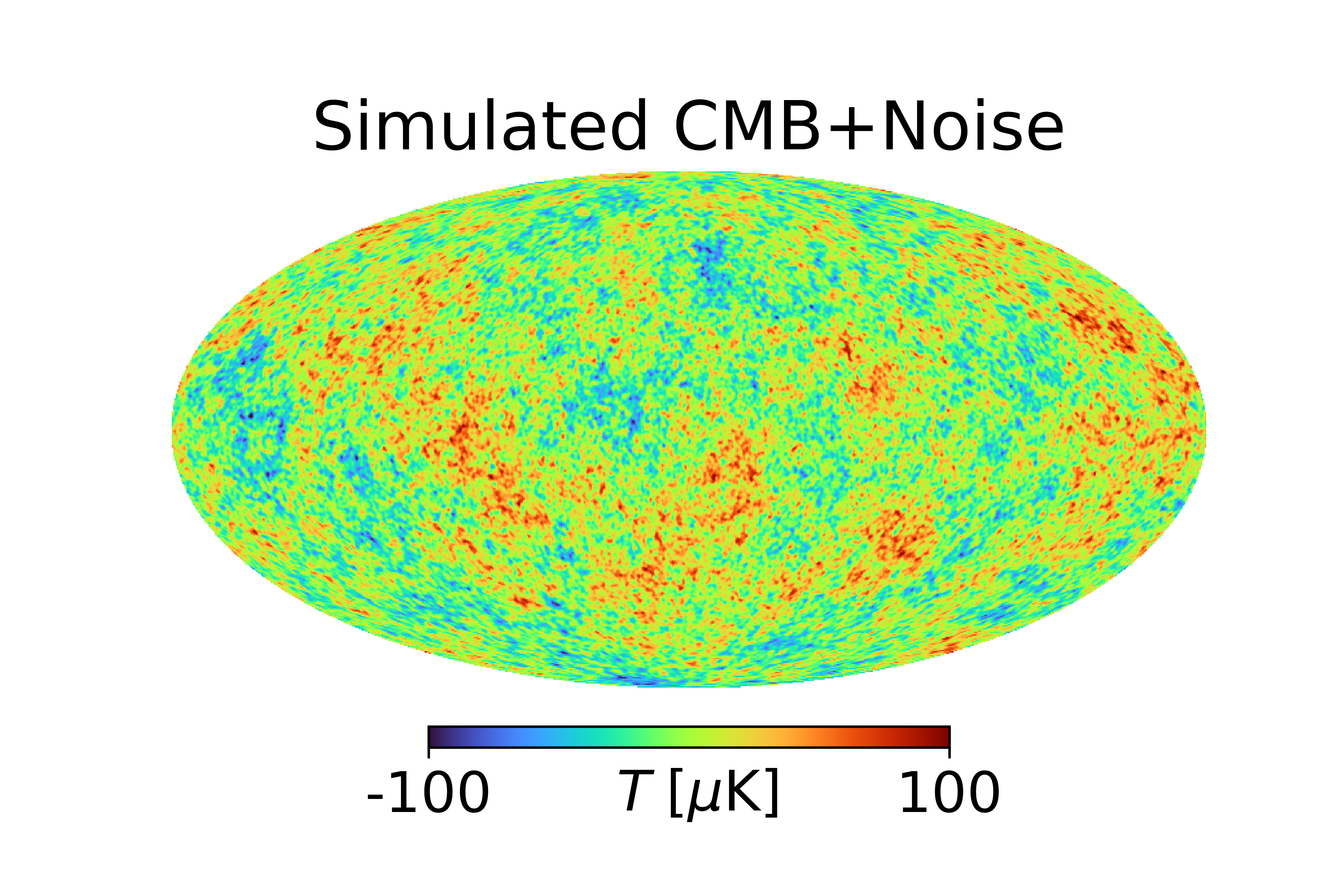}\\
 \hspace{-.5cm}\includegraphics[scale=.4]
 {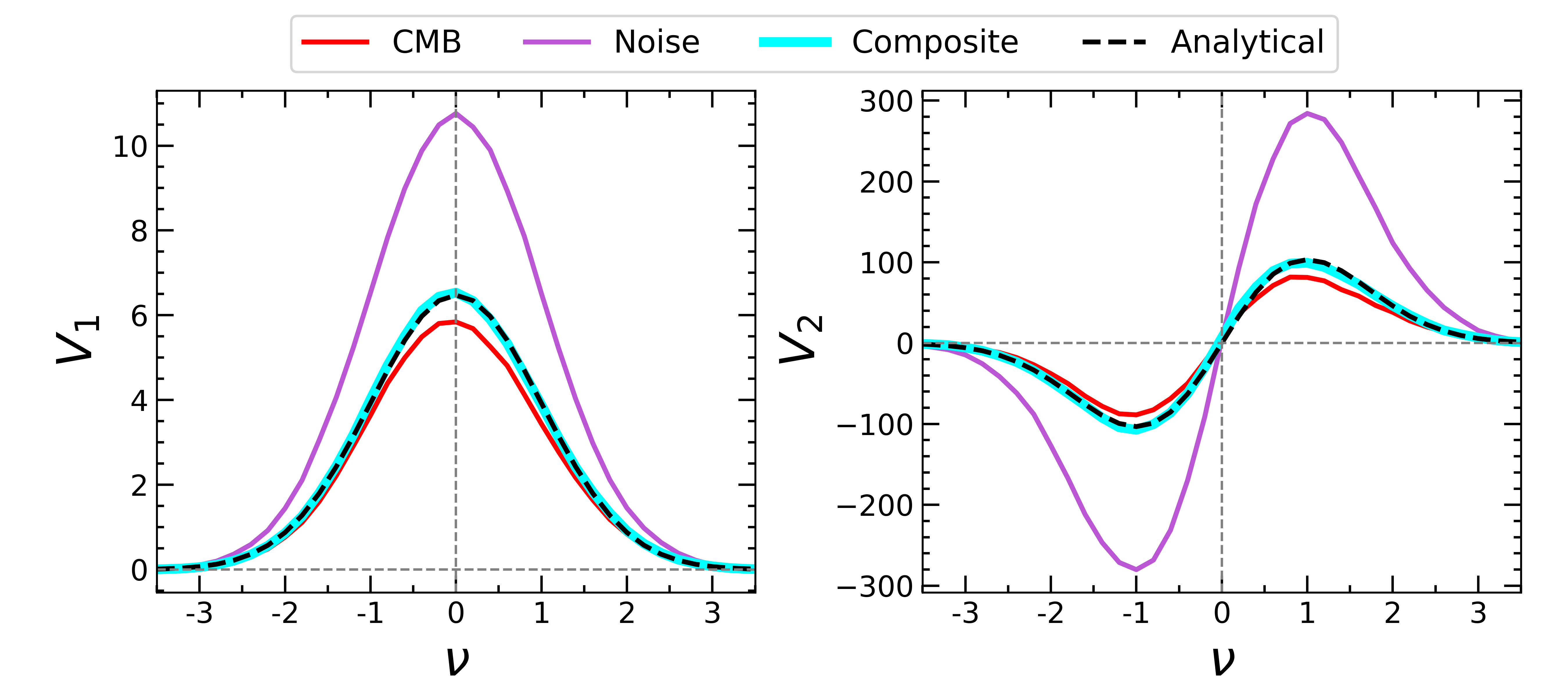}
\end{center}
\vspace{-0.6cm}
\caption{{\em Top}: Simulated Gaussian CMB (left) and noise (right) maps. {\em Middle}: The composite of CMB and noise maps. {\em Bottom}: The MFs $V_1$ (left) and $V_2$ (right)  for CMB SMICA (red), noise (purple) and their composite (cyan) maps. The plot for the analytic formula is shown by the black dashed lines which overlap the cyan lines.}
\label{fig:uvG}
\end{figure}

Let us consider $u$ to be a simulated Gaussian CMB map with input CMB power spectrum obtained from CAMB~\cite{Lewis:2011}. We use cosmological parameter values given by Planck~\citep{Planck2018I}. The map resolution is set by the Healpix~\cite{Gorski:2005} parameter $N_{\rm side}=256$. Let the secondary field $v$ be a toy example of a Gaussian noise map. We simulate $v$ at the same resolution as the CMB map, by assuming a power spectrum of the form $C_{\ell} \propto \ell$. This power spectrum is chosen so as to mimic the behaviour of noise in CMB experiments which have higher power towards high $\ell$ (small scales). We use a single map of each field. The simulated CMB  and noise maps are shown in the top panels of Fig.~\ref{fig:uvG}, while their composite map is shown in the middle panel. In this example we have $\e=0.4$ and $p=1.35$. It is difficult to distinguish the composite map from the CMB map by eye since $\e<1$.  

Since the sum of two uncorrelated Gaussian random variables is also a Gaussian variable, the composite field $f=u+v$ is also Gaussian. So, the presence of $v$ will modify only the amplitude of the MFs $V_1$ and $V_2$ of $f$ compared to $u$, while $V_0$ will not be affected.

 The bottom panels of Fig.~\ref{fig:uvG} show the numerically computed $V_1$ and $V_2$ versus $\nu$ for the  CMB (red), noise (purple) and their composite (cyan) maps.  $V_0$ is not shown since it is identical for all Gaussian fields.  
 The black dashed lines (which overlap the cyan lines) are plots of the Gaussian analytic formulae of the MFs with amplitude $A_k^f$. 
 We see good agreement of the analytic formulae with the numerically computed MFs.  From the discussion in section \ref{sec:sec3.1}, if $p>1$ the amplitudes of both $V_1$ and $V_2$ for the composite fields will be larger than that of the signal $u$, and this is what is obtained. 
 
These plots quantify the bias introduced by $v$ on the amplitude of the MFs. 
The bias is positive ($A_k^f > A_k^u$) for $p>1$, and negative ($A_k^f< A_k^u$) for $p<1$. We note that even though we have  used CMB and noise maps for this example, this result is applicable to any composite field. 


\subsection{Sum of mildly non-Gaussian CMB of local $f_{\rm NL}$ type and Gaussian noise maps}
\label{sec:sec4.2}

Let us consider the signal field $u$ to be a CMB temperature map with input local type $f_{\rm NL}$ non-Gaussianity~\cite{Salopek:1990,Gangui:1994,Komatsu:2001,Maldacena:2003JHEP}. In this case the leading contribution to the non-Gaussian deviations of the MFs will be given by $v_k^{(1)}$. The analytic expressions for the non-Gaussian deviations of the three MFs of $u$, given by Eq.~\ref{eqn:DVk_ana}, now become,
\bea
\D V_{0}^{{\rm ana},u}(\nu)&=& A_0\, e^{-\nu^2/2} \,\frac{S^{(0) u} }{6}H_2(\nu) \,\s^u_0,  \label{eqn:DV0_ana} \\
\D V_{1}^{{\rm ana},u}(\nu)&=& A_1^u\, e^{-\nu^2/2}  \left[ \frac{S^{(0) u}}{6}H_{3}(\nu) \right.\nn \\ && \left. \hskip 1.6cm +\frac{S^{(1)u}}{3}H_{1}(\nu) \right]\s^u_0,\nn\\  \label{eqn:DV1_ana}\\
\D V_{2}^{{\rm ana},u}(\nu)&=& A_2^u  \,e^{-\nu^{2}/2}  \left[ \frac{S^{(0)u}}{6}H_{4}(\nu) +\frac{2S^{(1)u}}{3}H_{2}(\nu) \right. \nn  \\ && \left.  
\hskip 1.6cm +\frac{S^{(2)u}}{3}H_{0}(\nu) \right] \s^u_0. \label{eqn:DV2_ana}
\eea

As an example of a composite field which is the sum of non-Gaussian and Gaussian fields, we take the second map, $v$, to be a Gaussian noise map, similar to the previous subsection. The analytic expressions for the non-Gaussian deviations of the three MFs of $f$ are given by Eqs.  \ref{eqn:DV0_ana} to \ref{eqn:DV2_ana}, but with two changes. $S^{(0)u}\s_0^u$, $S^{(1)u}\s_0^u$ and $S^{(2)u}\s_0^u$ will be replaced by  the corresponding expressions for $f$ given by   Eqs.  \ref{eqn:S0f} to \ref{eqn:S2f}. The amplitudes $A_1^u$ and $A_2^u$ will also be replaced by the corresponding ones for $f$.   

\begin{figure}[t]
\begin{center}
\includegraphics[scale=0.3]{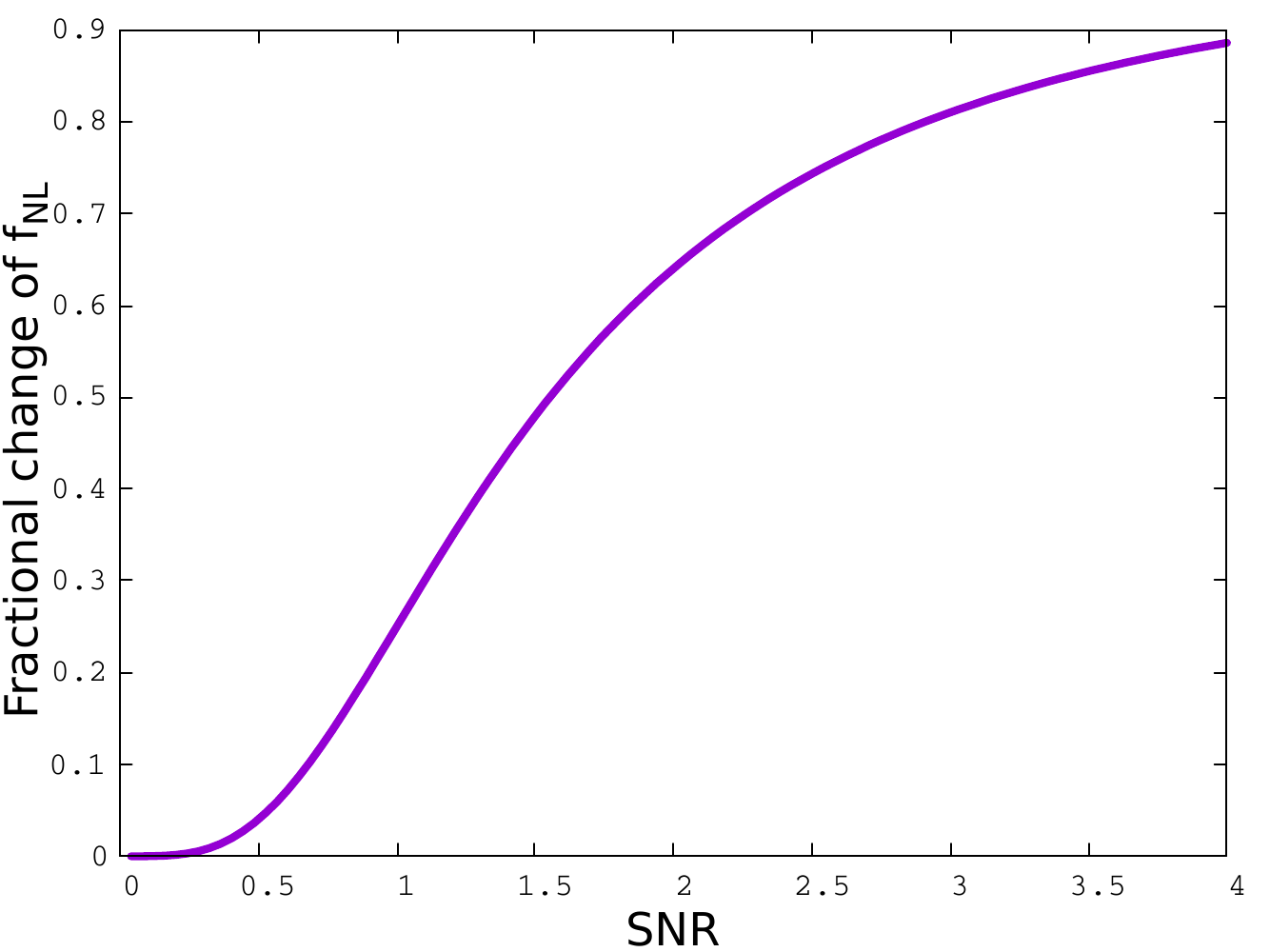}
\end{center}
\vspace{-0.3cm}
\caption{The fractional change of $f_{\rm NL}$, given by  $(\s^u/\s^f)^4 = 1/(1+\e^2)^2$, that is induced by the presence of Gaussian noise is shown as function of the SNR ($=\e^{-1})$.} 
\label{fig:df}
\end{figure}
 Before computing the MFs of $f$, it is instructive to discuss its PDF, which can be derived as follows. 
 It was shown in \cite{Ganesan:2015} that the PDF of a mildly non-Gaussian zero-mean random variable $u$ of local $f_{\rm NL}$ type is given by the form,
 \bea
 P_u(x) &=& \frac{1}{\sqrt{2\pi (\sigma_0^u)^2}}
 \exp\left\{ -\frac{x^2}{2(\sigma_0^u)^2} \right\} \nn \\
 && \times \left[ 1 +  f_{\rm NL}\sigma^u \,\left\{\frac{ x^3}{(\s_0^u)^3}-\frac{3x}{\sigma_0^u} \right\} \right],  \label{eqn:Pu} 
 \eea
 where $x$ denotes values of $u$ in its domain.  In terms of $\nu=u/\s_0^u$ we get
 \bea
 P_u(\nu)
 &=& \frac{1}{\sqrt{2\pi (\sigma^u)^2}}
 e^{-\nu^2/2}\left[ 1 +  f_{\rm NL}\sigma^u \,\bigg\{  \nu^3-3\nu \bigg\} \right]. \nn\\ \label{eqn:Punu}
 \eea
 The PDF of $v$ is
 \bea
 P_v(x) &=& \frac{1}{\sqrt{2\pi (\sigma^v)^2}} \exp\left\{ -\frac{x^2}{2(\sigma^v)^2} \right\}. \label{eqn:Pv}
 \eea
If $u$ and $v$ are uncorrelated, the PDF of $f$ 
is given by the convolution of their PDFs, 
 from which we get 
 \bea
 \hskip -.3cm P_f(x)  &=&  \frac{1}{\sqrt{2\pi (\sigma^f)^2}} \exp\left\{ -\frac{x^2}{2(\sigma^f)^2} \right\} \nn \\
 && \times \left[ 1 + f_{\rm NL} \s^u\,
 \left\{ \frac{(\s^u)^3 x^3}{(\s^f)^6} - 3 x \frac{(\s^u)^3}{(\s^f)^4} \right\} \right]. 
 \eea
In terms of $\nu=f/\s^f$, and redefining $f_{\rm NL}$ by  absorbing the factor  $\left(\s^u/\s^f\right)^4$, as
 \be
 \widetilde{f}_{\rm NL} = \left( \frac{\s^u}{\s^f}\right)^4   f_{\rm NL}, 
 \ee we get
 \bea
 P_f(\nu)  &=& \frac{1}{\sqrt{2\pi (\sigma^f)^2}}
 e^{-\nu^2/2} \left[ 1 +  \widetilde{f}_{\rm NL}\sigma^f  \,\bigg\{  \nu^3-3\nu \bigg\} \right]. \nn\\ \label{eqn:Pfnu}
 \eea

In terms of $\e$ we have $(\s^u/\s^f)^4 = 1/(1+\e^2 )^2$, where we have used the condition that $u,v$ are uncorrelated. Since $\s^u \le  \s^f$, we always have  $\widetilde{f}_{\rm NL} \le f_{\rm NL} $ and the fractional change is quantified by $\left(\s^u/\s^f\right)^4$.  
Comparing Eqs.~\ref{eqn:Punu} and \ref{eqn:Pfnu} we see that the functional form of the PDF of $f/\s^f$ is the same as that of $u/\s^u$, but with a decrease of the non-Gaussian level  that is   determined by the fractional change of $f_{\rm NL}$. 
In Fig.\,\ref{fig:df} we show a plot of the fractional change of $f_{\rm NL}$  versus the SNR ($=\e^{-1}$). At low SNR, the composite field is `Gaussianized'  due to the presence of Gaussian noise, and the true non-Gaussianity of the signal can be recovered only for ${\rm SNR}\gg 1$.

The area fraction $V_0$ is just the cumulative distribution function. Hence we can anticipate that the fractional change of $f_{\rm NL}$ will directly translate into decrease of $\D V_0$ for $f$ compared to $u$.  
However, the effect of the Gaussian noise on $\Delta V_1$ and $\Delta V_2$ cannot be inferred only from the PDF of $f$ because they encode the effect of the variable $p$ (first derivatives of the fields), and hence will not track the PDF directly.

\begin{figure*}[t]
\begin{center}
\includegraphics[scale=0.7]{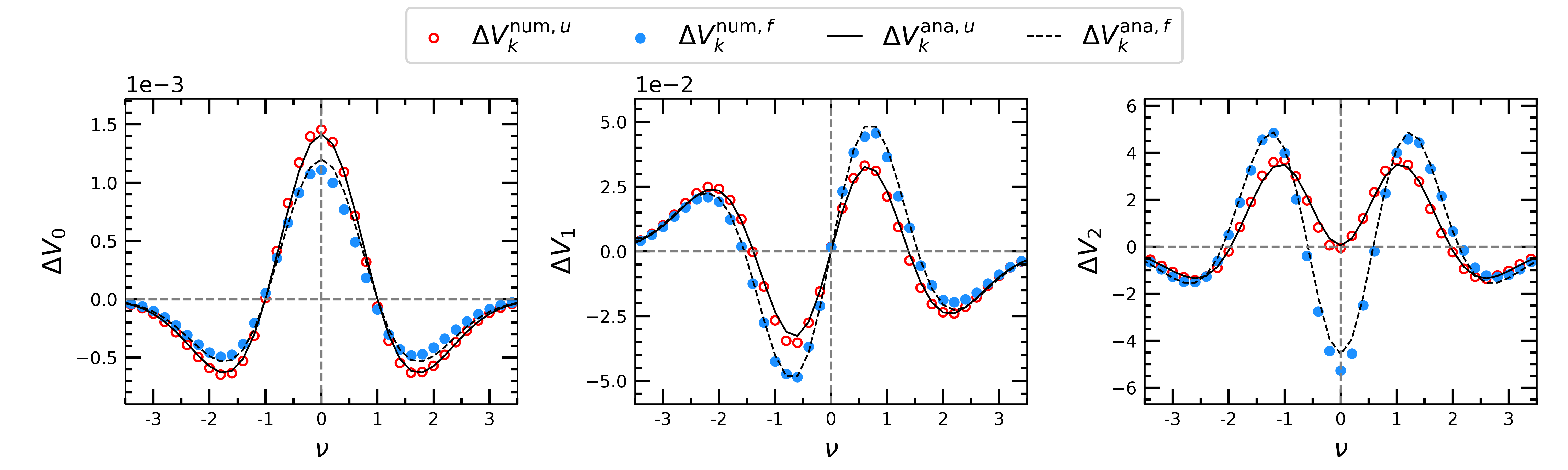}
\end{center}
\vspace{-0.7cm}
\caption{$\Delta V_k$ are shown for the non-Gaussian CMB temperature field $u$ (red hollow dots), and the composite field $f$ (blue solid dots). The corresponding $\Delta V_k^{\rm ana}$ are also shown for $u$ (black solid line), and for $f$ (black dashed line).}
\label{fig:dV_fnl}
\end{figure*}

We now focus on numerical computation of the non-Gaussian deviations of the MFs for $u$ and $f$, and comparison with the expected analytic expressions. For this purpose, let the non-Gaussian deviations of the numerically computed MFs (indicated by superscript `num') of a given field, whose nature is apriori unknown, be given by:
\begin{equation}{\label{eqn:DeltaVk}}
   \Delta V_{k}^{\rm num} =V_{k}^{\rm num} -V_{k}^{G}.
\end{equation}
Note that when we are given simulations of Gaussian and corresponding non-Gaussian (same seed) maps,  $\Delta V_k^{\rm num}$ can be obtained by computing the two terms on the right hand side of Eq.~\ref{eqn:DeltaVk} numerically from the respective maps. However, in practical situations we do not have Gaussian maps that correspond to a given observed map. In that case to get $V_k^{G}$ we can calculate $\s_0 $ and $\s_1$ for the given field, and use these values as inputs in the Gaussian analytic formulae given by Eq.~\ref{eqn:gmf}.  The numerically computed MFs  suffer from errors due to discretization of $\d$-function~\cite{Lim:2012} for identifying threshold boundaries and due to pixellization of the field. These errors are estimated from the Gaussian maps and subtracted from $V_{k}^{\rm num} $.

For the calculations in this subsection, we use Gaussian and corresponding local $f_{\rm NL}$ type non-Gaussian CMB temperature maps provided by Elsner \& Wandelt~\cite{Elsner:2009}. To obtain ensemble expectations we use 1000 maps. The maps are available with maximum multipole value $\ell_{\rm max}= 1024$. We use $N_{\rm side}=512$ and the maps are smoothed with FWHM=$30'$. The value of $f_{\rm NL}$ used for showing the results is 100. We use this relatively large value (the best fit value of $f_{\rm NL}$ from Planck CMB data is $-0.9\pm 5.1$ \cite{Planck:2018_iso}) so as to avoid statistical fluctuations because we have only 1000 maps with $\ell_{\rm max}=1024$.   
 From the simulated non-Gaussian CMB and noise maps we get $\bar \e \simeq 0.42$ and $\bar p\simeq 2.47$, where the overbars indicate that the values are average over the 1000 maps. 


Fig.~\ref{fig:dV_fnl} shows $\Delta V_k^{\rm num}$  for  $u$ (red hollow dots) and  $f$ (blue solid dots). Also shown are $\Delta V_k^{\rm ana}$ for $u$ (solid black lines) and $f$ (dashed black lines).  For $u$, we see good agreement between the numerical results and the analytic formulae,  as expected. This has been demonstrated in previous works (see e.g. \cite{Hikage:2006}). For $f$, we obtain good agreement between the numerical results and the analytic formulae that we have derived. The amplitude of $\Delta V_0^f$ is less than that of $u$, as anticipated from the PDF. The amplitudes of both  $\Delta V_1^f$ and  $\Delta V_2^f$ are found to increase, while their shapes are different from that of $u$.   
This is  due to the combined effects of  $A_k^f$ being larger than $A_k^u$ and the differing strengths of the generalized skewness cumulants.

\begin{figure}
\begin{center}
\includegraphics[scale=0.57]{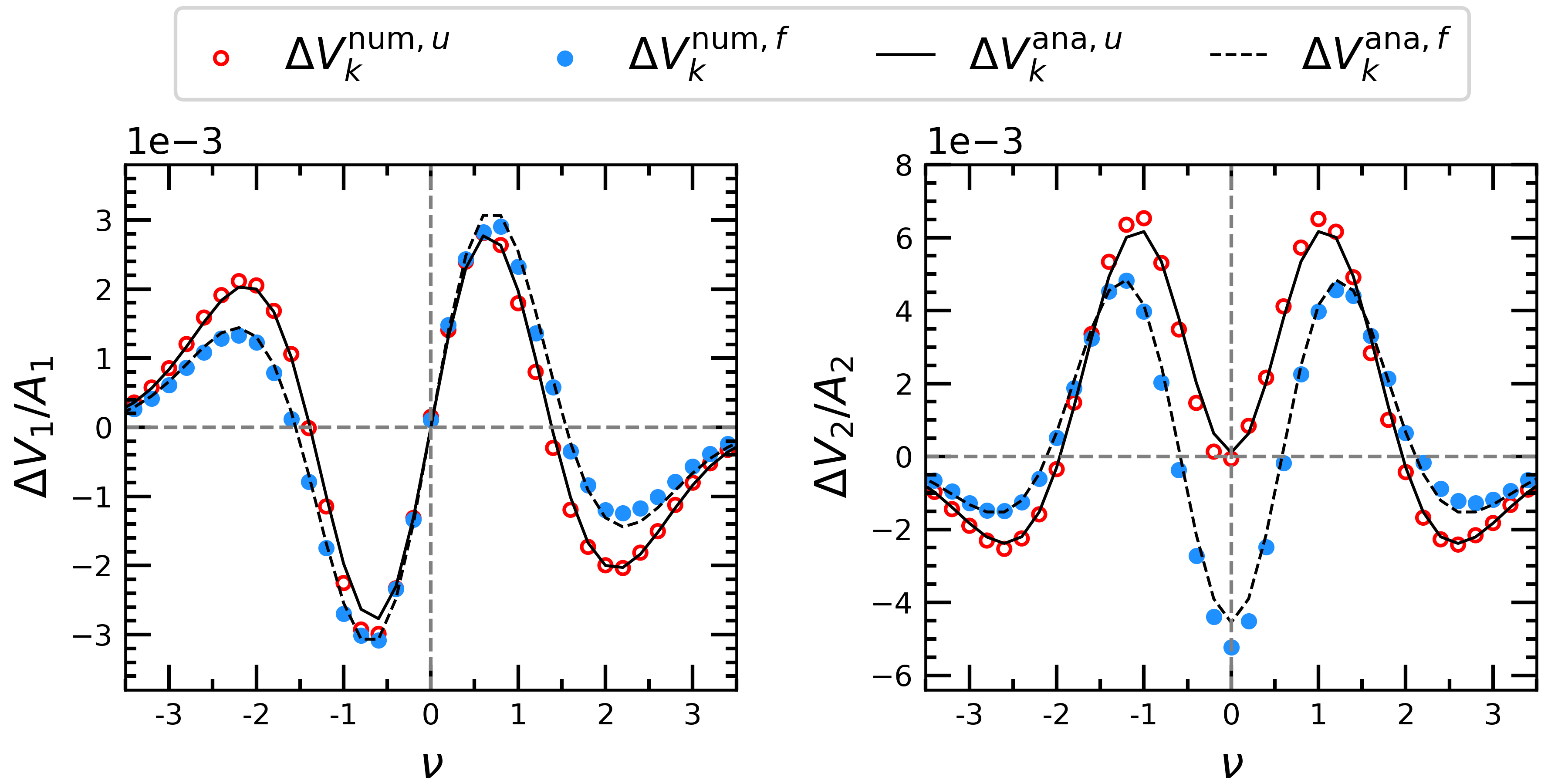}
\end{center}
\vspace{-0.6cm}
\caption{{\em Top}: $\Delta V_k/A_k$ are shown for  the non-Gaussian CMB temperature field $u$ (red hollow dots), and the composite field $f$ (blue solid dots). The corresponding $\Delta V_k^{\rm ana}/A_k$ are also shown for $u$ and $f$ by the black solid and dashed lines, respectively.}
\label{fig:dV_fnl2}
\end{figure}

To isolate the effect of noise  encoded in the generalized skewness cumulants from that on the amplitudes, we can scale out the amplitudes. In  Fig.~\ref{fig:dV_fnl2} we show $\Delta V_k/^{\rm num}A_k$ for  $u$ (red hollow dots) and $f$ (blue solid dots) for $k=1,2$. Again, the black lines show $\Delta V_k^{\rm ana}/A_k$ for $u$ (solid) and $f$ (dashed). We can see that the shapes for $f$ are different compared to the unscaled plots shown in Fig. \ref{fig:dV_fnl}.  As expected, there is good agreement between the numerical results and the analytic formulae.

\section{Conclusion}
\label{sec:sec5}
In this paper we have extended the analytic formulae for ensemble expectations of MFs for mildly non-Gaussian fields to composite fields which are sums of two fields. We derive the formulae explicitly for two cases - (i) both component fields are Gaussian and uncorrelated, (b) one field is mildly non-Gaussian while the second one is Gaussian. 
The formulae enable precise quantification of the effect of the secondary field on the morphology and statistical nature of the signal field, thereby providing analytic control over the calculations.   
In the context of cosmology, these formulae can be useful when dealing with observed cosmological data which are always  sums of true signals and secondary fields such as noise or residual contaminating signals.  

As concrete examples, the formulae are applied to two composite fields. The first is composed of Gaussian CMB temperature and Gaussian noise maps. The second one is composed of non-Gaussian CMB temperature of local $f_{\rm NL}$ type and Gaussian noise maps. 
In the first case we quantify the bias introduced by the presence of noise on the amplitudes of the MFs. The amount of bias  depends on the SNR  and the relative sizes of structures of the signal and noise fields. In the second example, apart from the amplitude bias, the presence  of noise introduces modification of the nature of non-Gaussianity of the composite field relative to that of the signal field. This modification can be quantified by determining the change of the shapes of the non-Gaussian deviations of the MFs of the composite field relative to the signal.

It is straightforward to extend the above explicit examples to cases where the secondary field is also mildly non-Gaussian.  Contamination of the CMB signal by residual Galactic foreground emissions is one such example~\cite{Chingangbam:2013}. It is also  straightforward, but tedious, to extend to cases where the two component fields are positively or negatively correlated. Examples of such cases are encountered when analysing different Galactic foreground emissions. Investigations of these examples will be taken up in the near future. We reiterate that the results of this paper are not confined to cosmological fields and can be applied to any physical example of composite fields.

\appendix
\section*{Acknowledgment}{We thank Stephen Appleby for a careful reading of the manuscript. We acknowledge the use of the \texttt{NOVA} HPC cluster at the Indian Institute of Astrophysics. Some of the results in this paper have been obtained by using the \texttt{CAMB}~\cite{Lewis:2011} and \texttt{HEALPIX}~\cite{Gorski:2005,Zonca:healpy,Healpix} software packages. Some plots are prepared using the \texttt{Matplotlib} package~\cite{Hunter:2007} and \texttt{Mathematica}.}

\def\apj{ApJ}%
\def\mnras{MNRAS}%
\def\aap{A\&A}%
\def\apjl{ApJ}
\def\aj{AJ}
\def\physrep{PhR}
\def\apjs{ApJS}
\def\jcap{JCAP}
\def\pasa{PASA}
\def\pasj{PASJ}
\def\nat{Natur}
\def\apss{Ap\&SS}
\def\araa{ARA\&A}
\def\aaps{A\&AS}
\def\ssr{Space Sci. Rev.}
\def\pasp{PASP}
\def\na{New A}
\def\prd{PRD}
\def\mathann{Math. Ann.}

\bibliographystyle{apsrev}
\bibliography{reference}

\end{document}